\documentclass[11pt]{article}
\usepackage{amssymb}
\usepackage{graphics}
\usepackage{epsfig}
\usepackage{a4wide}

\begin{document}
\newcommand{\eewwbb}{$e^+e^- \to W^+ W^- b\bar b$ }

\title{One-loop QCD corrections to the $e^+e^- \to W^+ W^- b\bar b$ process at the ILC}
\author{ Guo Lei, Ma Wen-Gan, Zhang Ren-You, and Wang Shao-Ming  \\
{\small Department of Modern Physics, University of Science and Technology}\\
{\small of China (USTC), Hefei, Anhui 230027, P.R.China}  }

\date{}
\maketitle \vskip 15mm
\begin{abstract}
We study the full contributions at the leading order(LO) and QCD
next-to-leading order(NLO) to the cross section of the \eewwbb
process in the standard model(SM) at the ILC. In dealing the
resonance problem we adopted the complex mass scheme in both
tree-level and one-loop level perturbative calculations. Our
numerical results show that the K-factor varies from $1.501$ to
$0.847$ when $\sqrt{s}$ goes up from $360~GeV$ to $1.5~TeV$. We
investigate the dependence of the LO and QCD NLO corrected cross
sections of process \eewwbb on colliding energy $\sqrt{s}$ and
Higgs-boson mass. We also present the results of the LO and QCD
NLO corrected distributions of the transverse momenta of final
particles, and the invariant masses of $Wb$-, $b\bar b$- and
$WW$-pair.
\end{abstract}

\vskip 5cm {\large\bf PACS: 13.66.Jn, 14.65.Ha, 14.80.Bn, 12.38.Bx
}

\vfill \eject

\baselineskip=0.32in

\renewcommand{\theequation}{\arabic{section}.\arabic{equation}}
\renewcommand{\thesection}{\Roman{section}.}
\newcommand{\nb}{\nonumber}

\newcommand{\Dir}{\kern -6.4pt\Big{/}}
\newcommand{\Dirin}{\kern -10.4pt\Big{/}\kern 4.4pt}
\newcommand{\DDir}{\kern -7.6pt\Big{/}}
\newcommand{\DGir}{\kern -6.0pt\Big{/}}

\makeatletter      
\@addtoreset{equation}{section}
\makeatother       

\section{Introduction}
\par
The Higgs boson, which gives masses to the weak vector bosons and
fermions, plays an important role in the standard model(SM).
Unfortunately, it has not been directly detected yet in
experiments. Searching for Higgs boson within the standard
model(SM) and study the phenomenology concerning Higgs properties
are the important tasks at the present and upcoming high energy
colliders. LEP II experiments have provided the lower limit on the
SM Higgs mass as $114.4~ GeV$ at the $95\%$ confidence level,
which is extracted from the results of searches for $e^+e^- \to
Z^0H^0$ production\cite{lower mH,upper mH}. While the indirect
evidences of the SM Higgs mass through electroweak precision
measurements indicate the $95\%$ C.L. upper bound as $m_H \lesssim
182~GeV$, when the lower limit on $m_H$ is used in determination
of this upper limit\cite{upper mH}. On the other hand, the heavy
top-quark practically plays a central and crucial role in probing
the electroweak symmetry breaking as well as the flavor problem in
all the extended models beyond the SM which address the hierarchy
problem. Recently, a new datum of top-quark mass has been already
presented by the CDF and D0 experiments at Fermilab, and the
preliminary world average mass of the top-quark is known as
$m_t=172.5 \pm 1.3(stat) \pm 1.9(syst)~GeV$, which corresponds to
a $20\%$ precision improvement relative to the previous
combination\cite{tew}.

\par
The future International Linear Collider (ILC) is proposed by the
particle physics community with the entire colliding energy in the
range of $200~GeV <\sqrt{s}<500~GeV$ and an integrated luminosity
of around $500~(fb)^{-1}$ in four years. The machine should be
upgradeable to $\sqrt{s}\sim 1~TeV$ with an integrated luminosity
of $1~(ab)^{-1}$ in three years\cite{ILC}. Most of the main
physics topics within the SM or its beyond at $TeV$ energy scale
can be explored at such a machine. Emphasis is given to the study
of top-quark physics, electroweak physics in the SM, and the
measurements in the extended SM, such as supersymmetry.

\par
Compared with the hadron colliders, such as the Tevatron and the
CERN Large Hadron Collider (LHC), the ILC can produce top and
Higgs boson signal events more easily resolved from backgrounds.
Therefore, the ILC is an ideal facility to study top and Higgs
physics with much more precise measurement for their parameters.
At the ILC we can also carry out the study of gauge boson
interactions, and the delicate cancellations which are related to
the gauge structure of the theory and essential to preserve
unitarity. Furthermore, the ILC experiment might be able to
explore the signature of the new physics, if the SM is really only
an effective theory at low energy.

\par
At the ILC, detecting the top-quark pair production process
$e^+e^- \to t \bar t$ is a good way to study the top-quark
properties, and the associated Higgs production with $Z^0$ boson
$e^+e^- \to H^0Z^0$ is one of the cleanest signature in
discovering Higgs boson if the the b-quark trigger system has high
performances except vertex detectors\cite{LEPlu}. The former
process will be followed by the subsequential decay through $t
\bar t \to W^+W^-b\bar b$\cite{bigi}, while the later process goes
via $H^0Z^0 \to W^+W^-b\bar b$ through decays $H^0 \to W^+W^-$ and
$Z^0 \to b\bar b$ if the Higgs boson mass is larger than
$2m_W$\cite{ball2}. Therefore, the signature of \eewwbb at the ILC
serves as non-resonant background to both top-quark pair
production and associated production of Higgs boson with $Z^0$
boson. We can see that it is crucial to separate the top and Higgs
signatures from the other $W^+W^-b\bar b$ production backgrounds
in ILC experimental data analyzing. In the precise measurements of
the signals of both the $t \bar t$ pair and $H^0Z^0$ associated
production processes, the relevant irreducible background from
\eewwbb should be carefully investigated.

\par
In Refs.\cite{bigi,widthtop,qcdcorr,softphoton, hardphoton,ewcorr}
the NLO electroweak and QCD corrections to the process $e^+e^- \to
t \bar t$ and decay $t \to W^+ b$ have been already extensively
studied. And the non-relativistic effect near the threshold of
$t\bar t$ production is also studied carefully in
Ref.\cite{topthreshold}, which can not be reliably described with
fixed QCD orders in perturbative theory. The Higgs-strahlung
Bjorken process $e^+e^- \to H^0Z^0$ was investigated in
Ref.\cite{ZH}, and the process $e^+e^- \to t\bar t \to W^+W^-b\bar
b \to 6f$ with six fermion final states after $W$ pair decays has
been also calculated at the lowest order in Ref.\cite{6f}. The
evaluation of the \eewwbb process with finite width method at the
tree-level is also presented in Ref. \cite{ball2, ball1}. All
those studies indicate that the precise investigations of the
characteristics of top-quark and the Higgs-boson are significant
for the future $e^+e^-$ ILC experiments.

\par
In this paper we present the calculations of the cross section of
the process \eewwbb at the leading order(LO) and its QCD
next-to-leading order(NLO) (${\cal O}(\alpha_s)$) corrections. The
paper is organized as follows: In the following section we present
the analytical calculations for process \eewwbb at the LO and QCD
NLO. The verifications of the correctness of our calculations are
declared in section III. The numerical results and discussions are
given in section IV. In the last section we give a short summary.

\vskip 5mm
\section{Calculations}
\par
The calculations for the process \eewwbb are carried out in 't
Hooft-Feynman gauge. In the QCD NLO calculations, we use the
dimensional regularization(DR) method to isolate the
ultraviolet(UV) and infrared(IR) singularities. In order to
preserve gauge invariance, we adopt the approach of the complex
mass scheme to deal with the unstable particles in the
calculations for the tree-level cross section and QCD NLO
radiative correction\cite{cms,complexmass}. The on-mass-shell(OS)
scheme is used to renormalize the masses and fields of related
bosons and fermions. The FeynArts3.2 package\cite{fey} is adopted
to generate Feynman diagrams and convert them to corresponding
amplitudes. The amplitude calculations are mainly implemented by
applying FormCalc4.1 programs\cite{formloop}. The formula for
calculating the IR divergent integrals with complex internal
masses in DR scheme are obtained by analytically extending the
expressions in Ref.\cite{Stefan} to the complex plane. The
numerical evaluations of IR safe one-point, two-point, three-point
and four-point integrals with internal complex masses, are
implemented by using the expressions analytically continued to
complex plane from those presented in
Refs.\cite{OneTwoThree,Four}. And the 5-point scalar integral can
be expressed in terms of multiple scalar four-point
integrals\cite{Five}. The subroutines for one-loop integrals with
complex masses are coded based on the LoopTools2.1\cite{formloop}
package which comes from FF library\cite{ff}. The $2\to 4$
phase-space integration routine\cite{guolei} is created based on
the 2to3.F program in FormCalc4.1 package. The five-body
phase-space integration for hard gluon radiation process
\eewwbb$g$ is accomplished by using CompHEP-4.4p3
program\cite{CompHEP}.

\par
\begin{figure*}
\begin{center}
\includegraphics[width=10.5cm]{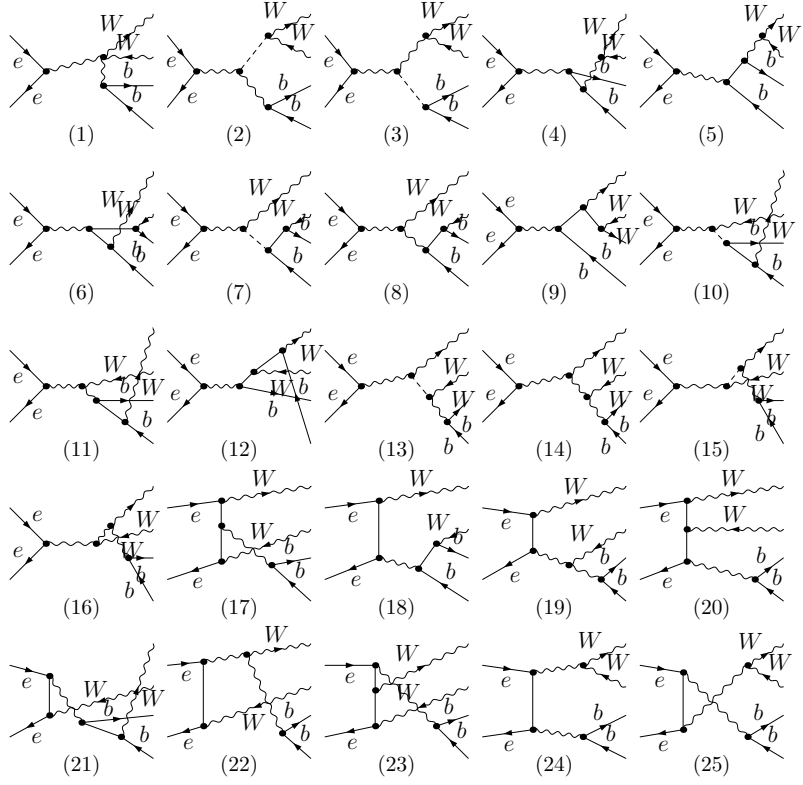}
\caption{\label{fig1} The generic tree-level Feynman diagrams for
the \eewwbb process. Internal wavy-line represents $\gamma$-,
$Z^0$-, or $W^\pm$-propagator. Internal dash-line represents a
Higgs boson $H^0$ or a Goldstone $G^0(G^{\pm})$.}
\end{center}
\end{figure*}

Now we present the analytically calculations of the tree-level
cross section for \eewwbb and its QCD NLO radiative corrections.
The notations for the process are defined as
\begin{equation}
\label{process} e^+(p_1)+e^-(p_2) \to W^+(p_3)+ W^-(p_4)+b(p_5)+\bar b(p_6),
\end{equation}
where $p_i~(i=1-6)$ label the four-momenta of incoming $e^+$,
$e^-$ and outgoing final particles, respectively. There are 64
generic tree-level diagrams for the process \eewwbb presented in
Fig.\ref{fig1}, where internal wavy-line represents $\gamma$,
$Z^0$, or $W^\pm$ and internal dash-line represents a Higgs-boson
$H^0$ or a Goldstone $G^0(G^{\pm})$. We can easily find that in
Fig.\ref{fig1} there includes the tree-level diagrams for the
processes $e^+e^- \to t^{*}\bar{t}^{*}\to W^+W^-b\bar{b}$ and
$e^+e^- \to H^{0*}Z^{0*} \to W^+W^-b\bar{b}$.
The differential cross section for the process \eewwbb at the
tree-level is obtained by the tree-level is obtained by
\begin{equation}
d\sigma_{tree}= \frac{(2 \pi)^4}{4\sqrt{(p_1\cdot
p_2)^2-m_e^4}}\overline{\sum}\left|{\cal M}_{tree}\right|^2
{d\Phi_4}, \label{Sig}
\end{equation}
where $d\Phi_4$ is the four-body phase space element given by
\begin{equation} \label{PSelement}
d\Phi_4  = \delta^ {(4)}\left (p_1+p_2-\sum_{i=3}^{6}p_i \right )\prod_{i=3}^{6}
\frac{d^3p_i}{(2\pi)^3 2E_i}.
\end{equation}
The summation in Eq.(\ref{Sig}) is taken over the spins and colors
of final states, and the bar over the summation recalls averaging
over initial spin states. In the calculation, the internal $Z^0$
and Higgs boson can be real, and the top-quark propagator can also
be resonance when $\sqrt{s}>2m_t$. To deal with these resonant
singularities, we use the so-called complex mass scheme(CMS) in
our perturbative calculations\cite{cms,complexmass}. The complex
masses of W-, Z-, H-boson and top-quark are defined as
\begin{equation}
\mu_X^2=m_X^2-im_X\Gamma_X, \qquad (X=W,Z,H,t).
\end{equation}
In the CMS approach the complex masses for all related unstable
particles should be taken everywhere in both tree-level and
one-loop level calculations. Then the gauge invariance can be
conserved and singularity poles of propagators are avoided.

\par
\begin{figure}[htbp]
\begin{center}
\includegraphics[width=11.5cm]{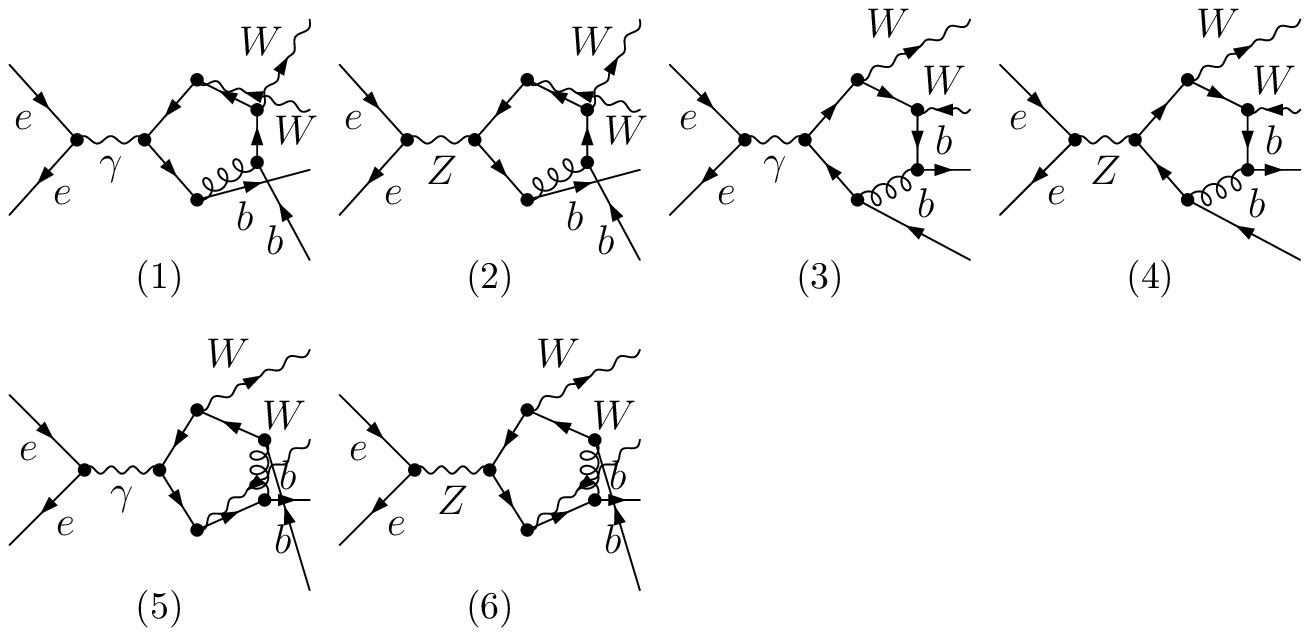}
\vspace*{-0.3cm} \centering \caption{\label{fig2} The pentagon
Feynman diagrams for the \eewwbb process. }
\end{center}
\end{figure}

In calculating the complete QCD NLO corrections, we should
consider the contributions of 30 self-energy diagrams, 94 triangle
diagrams, 17 box diagrams and 6 pentagon diagrams. As a
representative selection, we present the pentagon Feynman diagrams
of the \eewwbb process in Fig.\ref{fig2}. We adopt the Eqs.(4.26)
and (4.27) in Ref.\cite{complexmass} for the renormalized QCD
self-energy and counter-terms of top-quark with complex mass
neglecting terms of $O(\alpha_s^2)$ by using OS-scheme.

\par
There exist both ultraviolet(UV) divergency and infrared(IR) soft
singularity in the contributions of the QCD one-loop diagrams for
\eewwbb process, but no collinear IR singularity due to the
massive top- and bottom-quark. After doing the renormalization
procedure, the UV singularity is vanished.

\par
To cancel the IR soft divergency appeared in the virtual
correction, we should consider the contribution of the real gluon
emission process $e^+e^- \to W^+W^-b\bar{b}g$. We denote the real
gluon emission process as
\begin{eqnarray}
e^+(p_1)+e^-(p_2)\to W^+(p_3)+W^-(p_4)+b(p_5)+\bar b(p_6)+g(p_7).
\end{eqnarray}
To calculate the contribution of this process, we introduce an
arbitrary small soft cutoff $\delta_s$ to separate its 5-body
phase-space into two regions\cite{twocut}, i.e., soft($E_7 \leq
\delta_s\sqrt{s}/2$) and hard($E_7 > \delta_s\sqrt{s}/2$) regions.
After adopting the soft gluon approximation, the expression of
$\sigma_{soft}$ for $e^+e^- \to W^+W^-b\bar{b}g$ process with soft
gluon has the form as
\begin{equation}
d\sigma_{soft}=C_F\frac{\alpha_s}{2\pi}g_{56}~d\sigma_{tree},
\end{equation}
where $C_F=4/3$ and $g_{56}$ are defined as:
\begin{eqnarray}
g_{56} &=& \left(\frac{\pi\mu^2}{\Delta E^2}\right)^\epsilon
\Gamma(1+\epsilon)\, \left[\frac{4(p_5\cdot
p_6)}{\lambda^{1/2}(s_{56},m_b^2,m_b^2)}\ln(\sigma)+2\right]
\frac{1}{\epsilon}-\frac{2(p_5\cdot
p_6)}{\lambda^{1/2}(s_{56},m_b^2,m_b^2)} \nb \\ &&
 \times \left[\ln^2(\sigma)
+4Li_2\left(1-\sigma\right) \right] -\frac{2}{\rho}\ln\sigma
+{\cal O(\epsilon)}.    \label{g56}
\end{eqnarray}
In above equation, $\lambda(s_{56},m_b^2,m_b^2)$ is the kinematical
function defined by:
\begin{equation}\label{Lamb}
\lambda(x,y,z)=x^2+y^2+z^2-2xy-2yz-2zx,
\end{equation}
$\Delta E=E_7=\delta_s\sqrt{s}/2$,
$s_{56}=(p_5+p_6)^2$ and
\begin{equation}
\rho=\frac{\lambda^{1/2}(s_{56},m_b^2,m_b^2)}{s_{56}}, ~~~~
\sigma=\frac{1-\rho}{1+\rho}.
\end{equation}

\par
Our created $2\to 4$ phase space integration routine\cite{guolei},
is adopted in the tree-level and one-loop level calculations for
\eewwbb process. The IR singularity part of the soft gluon
emission process \eewwbb$(g)$ can be exactly cancelled by the IR
singularity induced by the one-loop virtual gluon correction. We
apply CompHEP-4.4p3 program\cite{CompHEP} to implement the phase
space integration of the hard gluon emission process \eewwbb$+g$.
Finally, we get the finite total cross section including complete
NLO QCD corrections for the process \eewwbb by summing up all the
contribution parts,
\begin{eqnarray}
\sigma_{NLO}=\sigma_{tree}+\sigma_{virtual}+\sigma_{soft}+\sigma_{hard}.
\end{eqnarray}

\vskip 5mm
\section{Checks}

\par
We have performed the following checks to prove the reliability of
our calculation:
\begin{itemize}
\item The LO cross section for the process \eewwbb is calculated
in the conditions of taking $\sqrt{s}=500~GeV$ and neglecting the
contribution of the diagrams with internal Higgs-boson exchange
which are taken in Ref.\cite{ball1}. The numerical results of the
LO cross section for the process \eewwbb are listed in Table
\ref{tabcheck}. There our results are obtained by using both
CompHEP-4.4p3 program and our created $2 \to 4$ phase-space
integration routine, and compared with the corresponding ones
presented in Ref.\cite{ball1}. We can see there is a good
agreement between ours and those presented in Ref.\cite{ball1}.
The in-house $2\to 4$ phase-space integration routine was also
once verified in our previous work\cite{guolei}.

\begin{table}
\begin{center}
\begin{tabular}{|c|c|c|c|}
\hline $m_t(GeV)$ & $\sigma_{LO}(fb)$(Ref.\cite{ball1})
& $\sigma_{LO}(fb)$(Comphep) & $\sigma_{LO}(fb)$ (ours)\\
\hline
150 & 663.11 & 663.03(1$\pm$0.05\%)  & 663.19(1$\pm$0.05\%)  \\
180 & 576.26 & 576.19(1$\pm$0.04\%)  & 576.52(1$\pm$0.04\%)  \\
200 & 497.63 & 497.58(1$\pm$0.04\%)  & 497.68(1$\pm$0.04\%)  \\
\hline
\end{tabular}
\end{center}
\begin{center}
\begin{minipage}{15cm}
\caption{\label{tabcheck}The comparison of the numerical results
of LO cross section neglecting the diagrams with Higgs-boson
interchanging by using CompHEP-4.4p3 system, our in-house $2\to 4$
phase-space integration routine with the corresponding selected
results presented in Ref.\cite{ball1} when $\sqrt{s}=500~GeV$. }
\end{minipage}
\end{center}
\end{table}

\item We use our created codes for numerical evaluation of the
one-loop integrals with complex internal masses. The comparisons
are made between the results and those obtained by doing directly
the integration of Feynman-parameter. There exists a good
agreement. The results from both calculations for scalar two-,
three-, four-point integrals are coincident with each other at
least up to six digits, respectively.

\item The exact cancellations of UV- and IR-divergencies are
verified both analytically and numerically in our calculation.

\item The independence of the total cross section including the
NLO QCD corrections on the soft cutoff $\delta_s(=2~\Delta
E/\sqrt{s})$ is confirmed numerically. Our calculation shows the
errors of the independence are less than $0.6\%$ in the $\delta_s$
region of $[10^{-4},~5\times 10^{-2}]$. In further numerical
calculation we fix $\delta_s=10^{-3}$.

\item In the following section, we shall clarify other verifications.
\end{itemize}

\vskip 5mm
\section{Numerical results and discussion}
\par
In our numerical calculation we take the following input
parameters\cite{hepdata,leger}:
\begin{equation}
\begin{array}{lll}
\alpha(m_Z)^{-1}=127.918, &\alpha_s(m_Z^2)=0.1176, &m_Z=91.1876~GeV, \\
m_W=80.403~GeV, &\Gamma_Z=2.495~GeV, &\Gamma_W=2.141~GeV, \\
m_t=172.5~GeV, &m_b=4.7~GeV,  &m_e=0.5109991~MeV.
\end{array}
\end{equation}
Due to the application of the CMS approach, we use the complex
weak mixing angle defined as
\begin{equation}
c_w^2=1-s_w^2=\frac{\mu_W^2}{\mu_Z^2}.
\end{equation}
In our LO and NLO numerical calculations we set the QCD
renormalization scale $\mu$ as $\mu=m_W+m_b$, and take the strong
coupling $\alpha_s(\mu^2)=0.11885$, which is obtained by using the
formula at three-loop level ($\overline{MS}$ scheme) with the five
active flavors\cite{hepdata}.

\par
Since the widths of top-quark and Higgs boson haven't been well
provided or measured experimentally by now, we use their
theoretical results from perturbative calculations. Considering
the fact that top-quark mass is above $m_W+m_b$, and $V_{tb}\sim
1$, the decay of top-quark is dominated by undergoing two-body
decay $t\to W^+b$, and the total decay width of top-quark is
approximately equal to the decay width of $t \to W^+b$. Neglecting
terms of order $m_b^2/m_t^2$, $\alpha_s^2$ and
$(\alpha_s/\pi)M_W^2/m_t^2$, the width predicted in the SM is
\cite{twidth}:
\begin{equation}
\Gamma_t=\frac{\alpha
m_t^3}{16m_W^2(1-m_W^2/m_Z^2)}\left(1-\frac{m_W^2}{m_t^2}\right)^2
\left(1+2\frac{m_W^2}{m_t^2}\right) \left[
1-\frac{2\alpha_s}{3\pi}\left(\frac{2\pi^2}{3}-\frac{5}{2}
\right)\right].
\end{equation}
The reasonable physical decay width of Higgs boson is obtained by
employing the program Hdecay\cite{hdecay}, where the partial decay
width $\Gamma (H^0 \to q\bar q)$ is calculated including ${\cal
O}(\alpha_s^3)$ QCD radiative corrections. Then we obtain
$\Gamma_t=1.3745~GeV$, $\Gamma_H(m_H=120~GeV)=0.3692\times
10^{-2}~GeV$ and $\Gamma_H(m_H=180~GeV)=0.6286~GeV$.

\par
\begin{figure}
\includegraphics[scale=0.36]{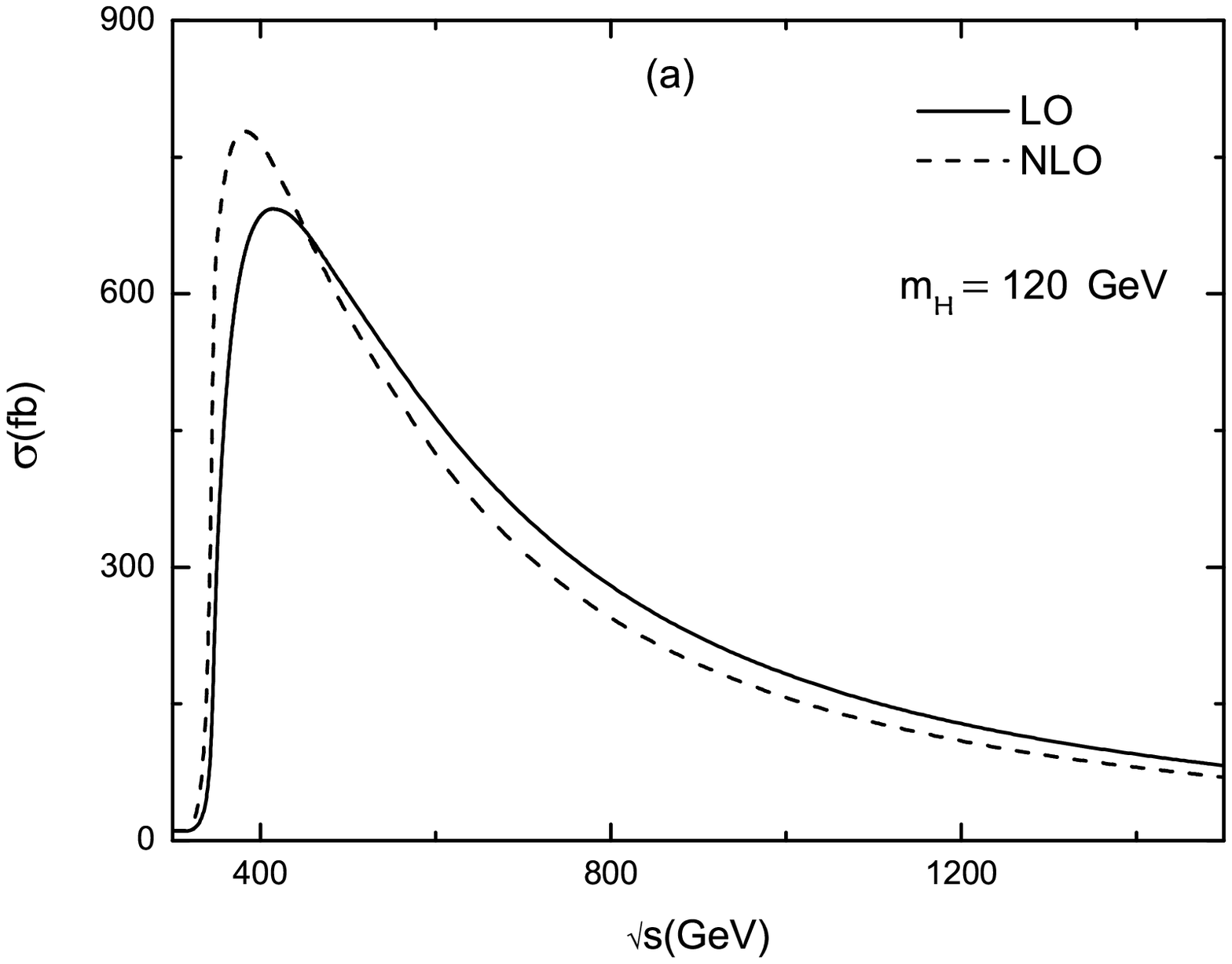}
\includegraphics[scale=0.36]{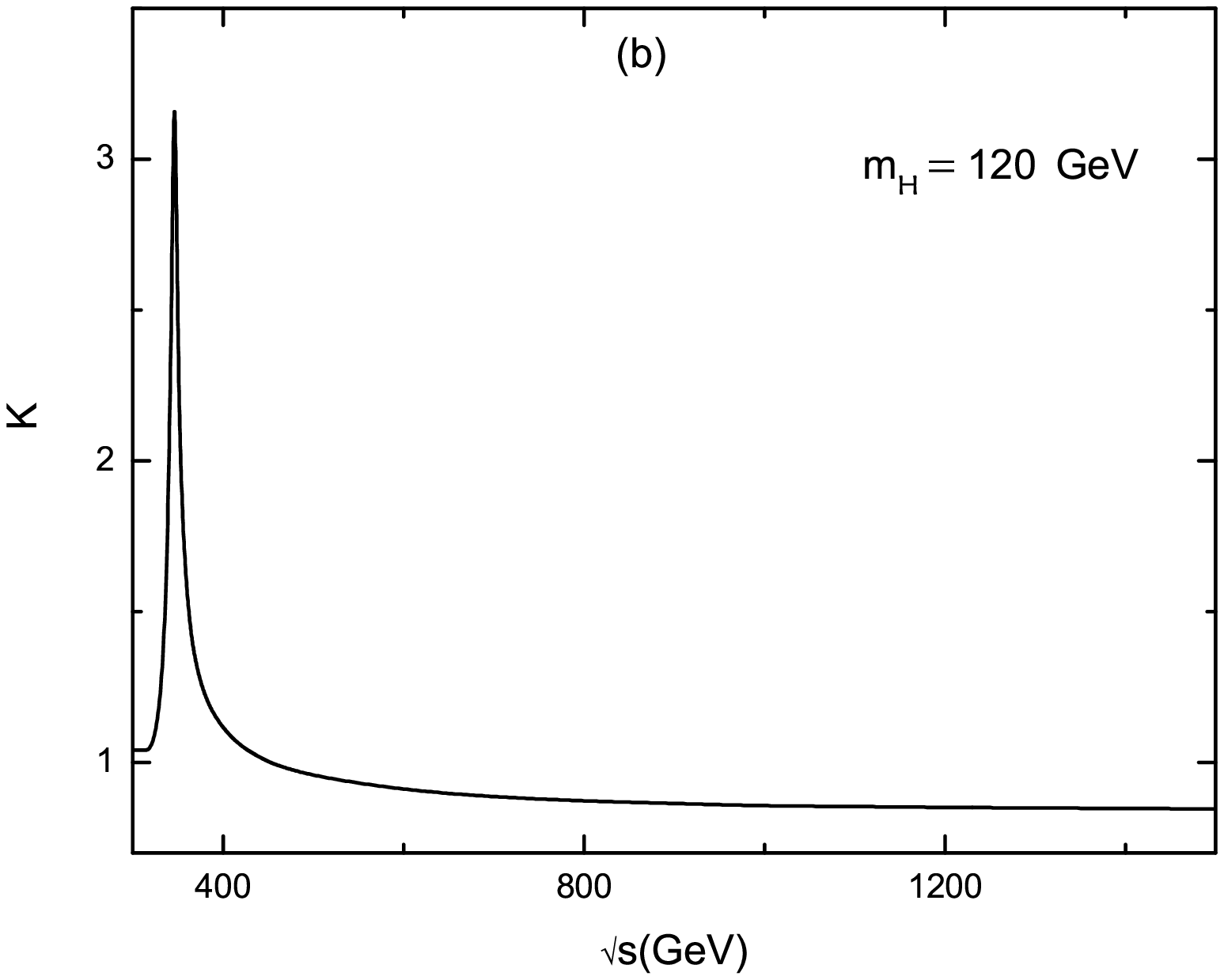}
\caption{\label{fig3} (a) The LO and NLO QCD corrected cross
sections for the process \eewwbb as the functions of c.m.s.
colliding energy($\sqrt{s}$) with $m_H=120~GeV$, (b) the
corresponding K-factor versus $\sqrt{s}$.}
\end{figure}

The numerical results of the LO, QCD NLO corrected cross sections
and the corresponding K-factor($K \equiv\frac{\sigma_{NLO}}
{\sigma_{LO}}$) for the process \eewwbb are plotted in
Figs.\ref{fig3}(a) and (b) respectively, when $m_H=120~GeV$. As
indicated in Fig.\ref{fig3}(a), both curves for the cross sections
at the LO and NLO increase quickly in the $\sqrt{s}$ region of
$[350~GeV,~400~GeV]$ and decrease when $\sqrt{s}>430~GeV$.
Fig.\ref{fig3}(b) shows that the corresponding K-factor decreases
slowly from $1.501$ to $0.847$ as $\sqrt{s}$ running from
$360~GeV$ to $1.5~TeV$. The large positive peak near the $t\bar t$
threshold in Fig.\ref{fig3}(b) is due to a Coulomb singularity
effect coming from the instantaneous gluon exchange between heavy
quarks which has a small spatial momentum. In Table \ref{tab1} we
list the values of $\sigma_{tree}$, $\sigma_{NLO}$ and K-factor at
some typical $\sqrt{s}$ points, which are read out from
Figs.\ref{fig3}(a-b). Since the QCD correction to the \eewwbb
process with high colliding energy can be approximately decomposed
into the QCD correction to the $t \bar t$ production plus the
corrections to the $t(\bar t) \to W^+ b(W^- \bar b)$ decays when
$m_H<2m_W$, we make following verification to check our results.
We evaluate the QCD correction to \eewwbb process by combining the
QCD corrections to $e^+e^- \to t\bar t$ production and $t(\bar t)
\to W^+ b(W^- \bar b)$ decays together, and get the K-factors to
process \eewwbb as $0.8562(1)$ for $\sqrt{s}=1~TeV$ and
$0.8433(1)$ for $\sqrt{s}=1.5~TeV$, which are coincident with the
corresponding ones in Table \ref{tab1} in error ranges.

\begin{table}
\begin{center}
\begin{tabular}{|c|c|c|c|}
\hline $\sqrt{s}(GeV)$ & $\sigma_{tree}(fb)$
& $\sigma_{NLO}(fb)$ & K-factor \\
\hline
500  & 602.57(1$\pm$0.05\%) & 575.5(1$\pm$ 0.38 \%) & 0.955(4) \\
1000 & 182.24(1$\pm$0.04\%) & 156.7(1$\pm$ 0.38 \%)  &  0.860(4) \\
1500 &  82.73(1$\pm$0.04\%) &  70.1(1$\pm$ 0.37 \%) &  0.847(4) \\
\hline
\end{tabular}
\end{center}
\begin{center}
\begin{minipage}{15cm}
\caption{\label{tab1} The LO and NLO QCD corrected cross sections,
K-factors for \eewwbb process with $m_H=120~GeV$ and
$\sqrt{s}=500~GeV$, $1000~GeV$, $1500~GeV$, respectively.}
\end{minipage}
\end{center}
\end{table}

\par
\begin{figure}
\centering
\includegraphics[scale=0.36]{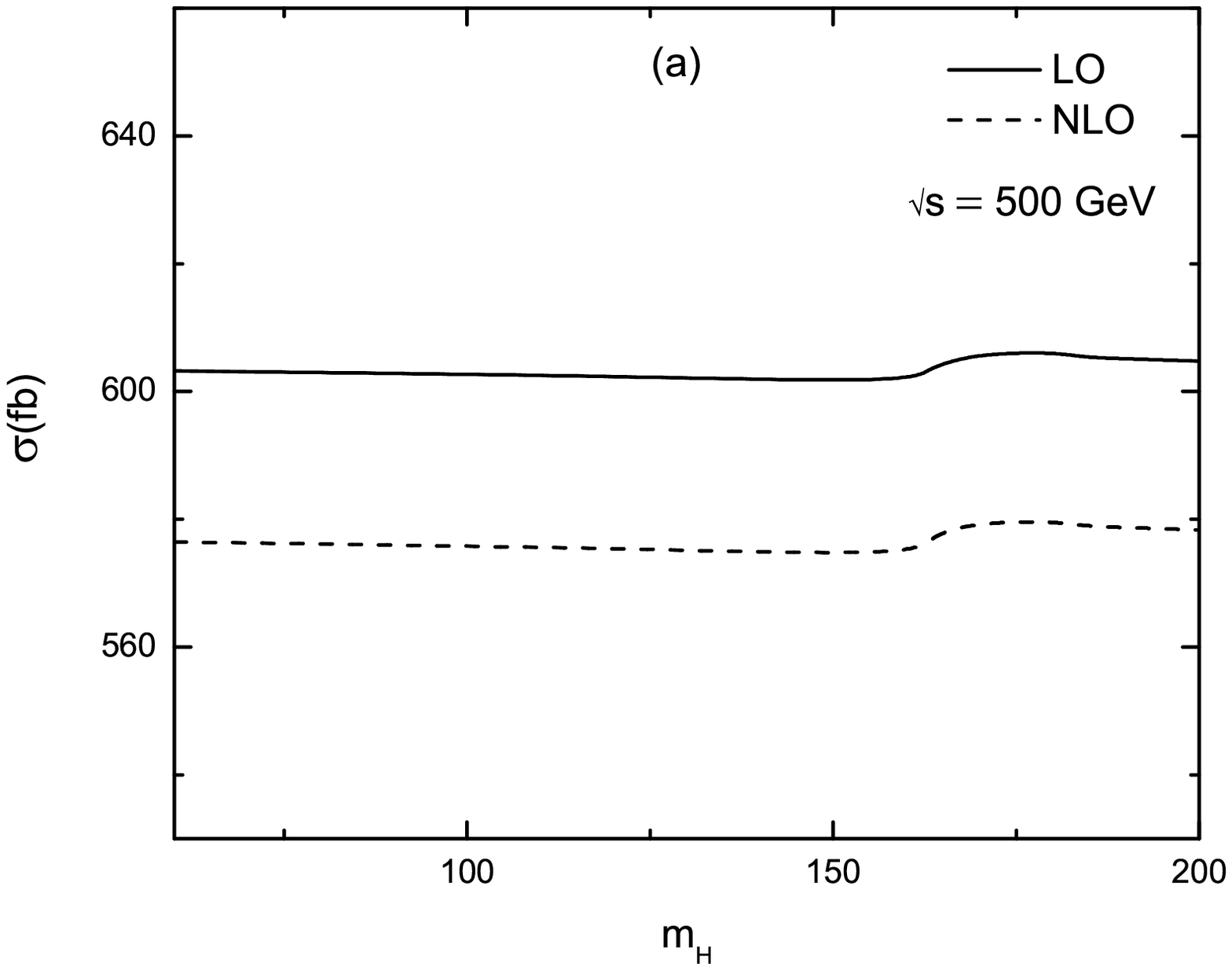}
\includegraphics[scale=0.36]{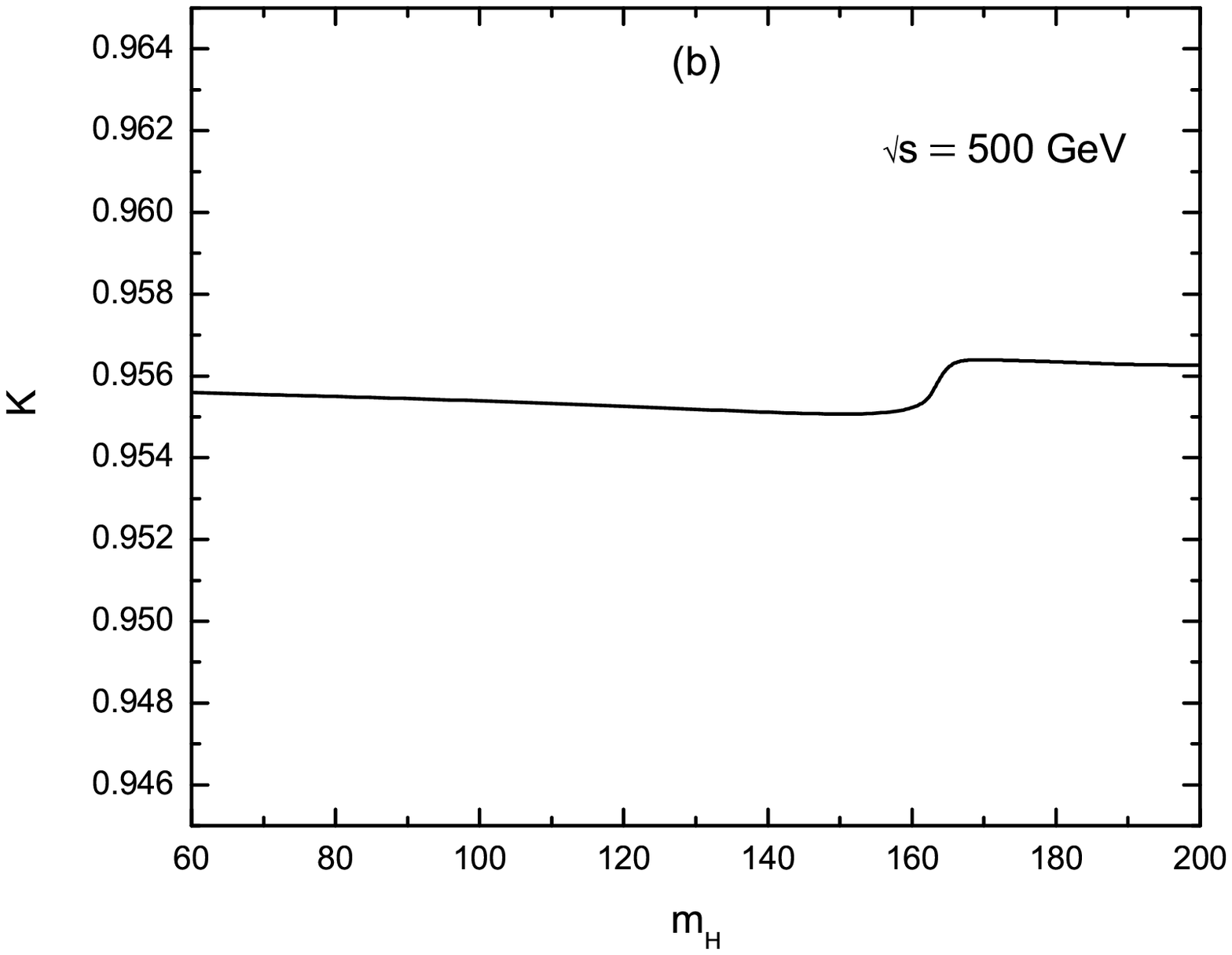}
\caption{\label{fig4} (a) The LO and QCD NLO corrected cross
sections for the process \eewwbb as the functions of Higgs
mass($m_H$) with $\sqrt{s}=500~GeV$. (b) The corresponding
relative QCD NLO corrections versus $m_H$.}
\end{figure}

In Fig.\ref{fig4}(a) we present the plot of the LO and QCD NLO
corrected cross sections as the functions of Higgs-boson mass,
with $\sqrt{s}=500~GeV$ and $m_H$ running form $60~GeV$ to
$200~GeV$. We find from Fig.\ref{fig4}(a) that the LO and QCD NLO
corrected cross sections are non-sensitive to the Higgs-boson mass
except in the vicinity of $m_H \sim 2 m_W$, from there the Higgs
mass becomes larger than $2m_W$, and $H^0$-, $Z^0$-boson are
simultaneously resonances. We can see also from the figures that
the contribution via $e^+e^- \to t^{*}\bar{t}^{*} \to
W^+W^-b\bar{b}$ channel is much larger than that from $e^+e^- \to
H^{0*}Z^{0*} \to W^+W^-b\bar{b}$ as concluded in Ref.\cite{ball1}.
Fig.\ref{fig4}(b) shows the corresponding K-factor has the values
around $0.956$ in the range of $m_H \in [60~GeV,~200~GeV]$.

\par
\begin{figure}
\centering
\includegraphics[scale=0.36]{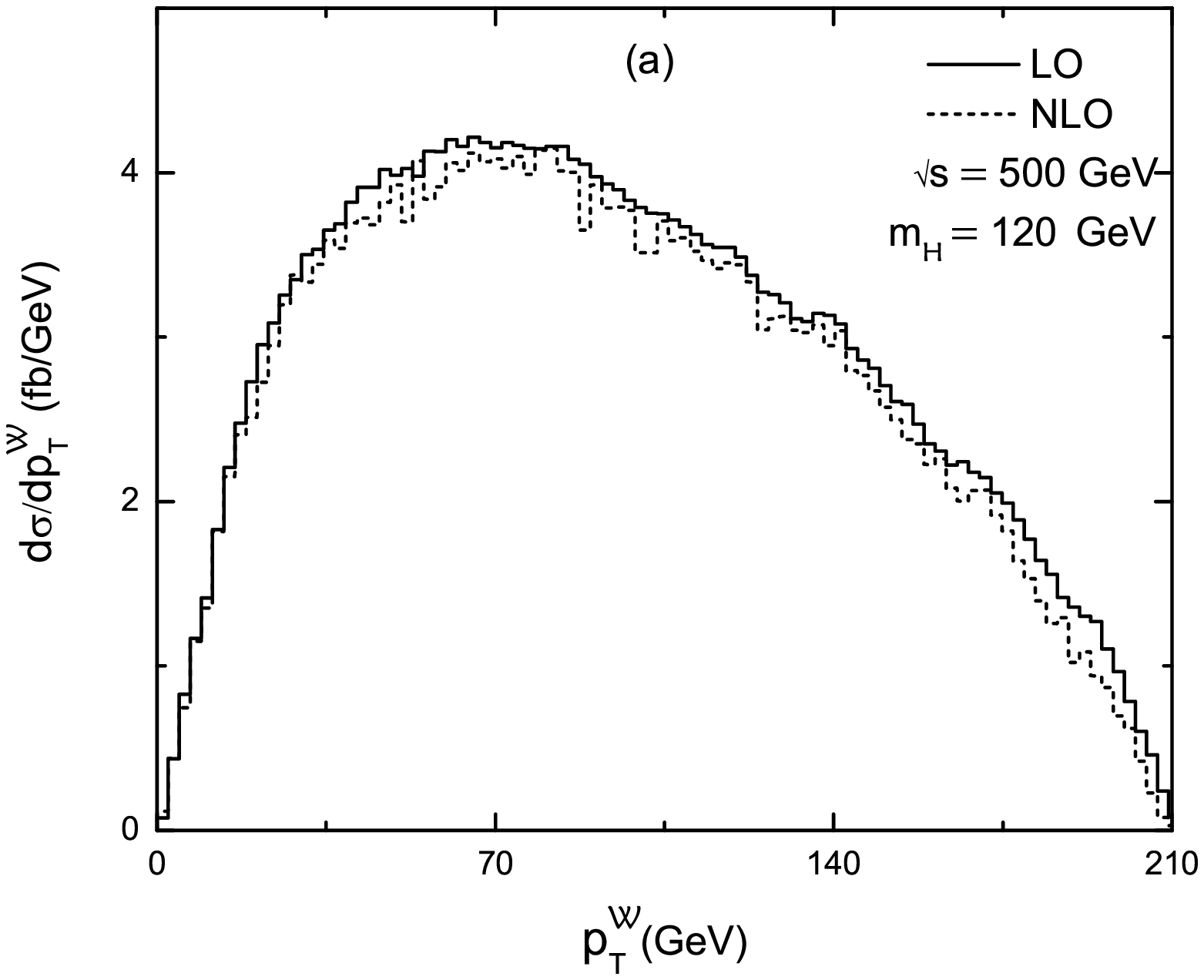}
\includegraphics[scale=0.36]{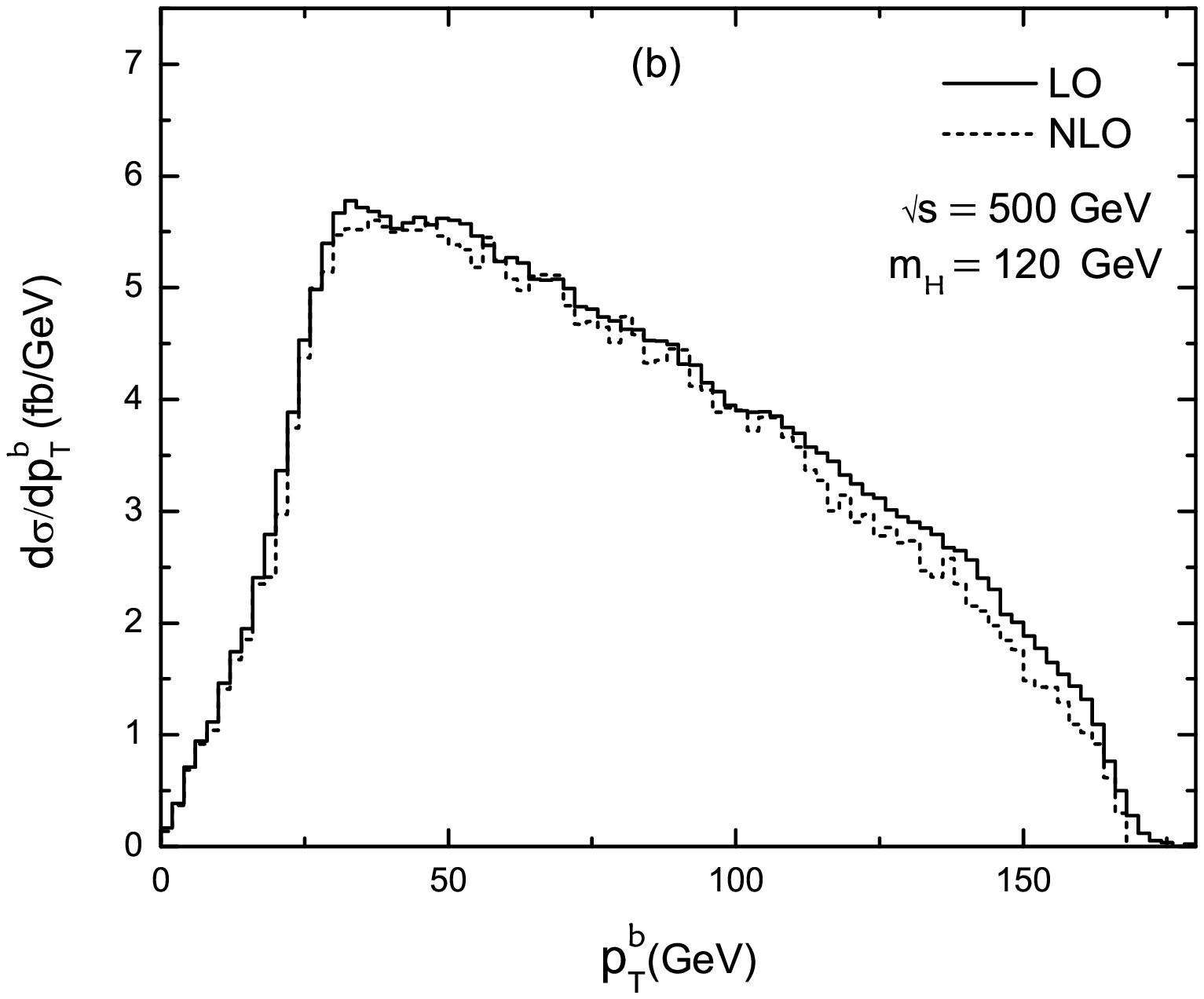}
\caption{\label{fig5} The distributions of the transverse momenta
of $W^+$ and bottom-quark for the \eewwbb process at the LO and
QCD NLO with $\sqrt{s}=500~GeV$ and $m_H=120~GeV$. (a) for $W^+$,
(b) for bottom-quark. }
\end{figure}

Due to the CP-conservation, the distributions of transverse
momenta of $W^-$-boson and $\bar b$ quark should be the same as
those of $p_T^{W^+}$ and $p_T^b$, respectively. We only present
the distributions of the $p_T^{W^+}$ and $p_T^b$ with
$m_H=120~GeV$ and $\sqrt{s}=500~GeV$ in Figs.\ref{fig5}(a) and
(b). In these two figures we can see that the QCD NLO corrections
suppress the LO differential cross sections
$d\sigma_{LO}/dp_T^{W^+}$ and $d\sigma_{LO}/dp_T^b$. They also
show that the differential cross sections of
$d\sigma_{LO,NLO}/dp_T^{W^+}$ and $d\sigma_{LO,NLO}/dp_T^b$ have
their maximal values at about $p_T^{W^+}\sim 70~GeV$ and
$p_T^b\sim 30~GeV$ respectively. We see that the line shapes of
the differential cross sections in these two figures are mainly
determined by the contributions of the of $e^+e^- \to
t^{*}\bar{t}^{*} \to W^+W^-b\bar b$ production.

\par
\begin{figure}
\centering
\includegraphics[scale=0.36]{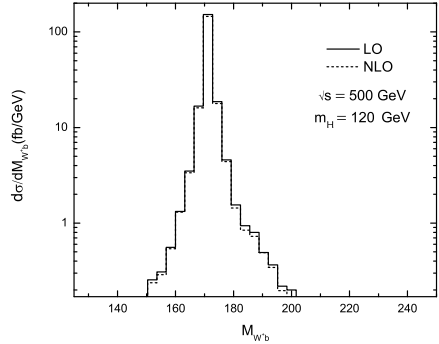}
\caption{\label{fig6} The distributions of the invariant mass of
$(W^+b)$-pair (or $(W^-\bar b)$-pair) at the LO and QCD NLO with
$m_H=120~GeV$ and $\sqrt{s}=500~GeV$. }
\end{figure}

We plot the invariant mass distributions of $(W^+b)$-pair, denoted
as $M_{(W^+b)}$, at the LO and QCD NLO in Fig.\ref{fig6} with
$m_H=120~GeV$ and $\sqrt{s}=500~GeV$. The distribution of
$M_{(W^-\bar b)}$ should be the same as that of $(W^+b)$-pair due
to the CP-conservation. We can see from the figure that most of
the events are concentrated around a peak located at the position
of $M_{(W^+b)}\sim m_t$. That demonstrates again the main
contribution to the cross section of the process \eewwbb with high
colliding energy, is from top-pair production channel $e^+e^-\to t
\bar t$ and followed by the decay of $t(\bar t) \to W^+b(W^-\bar
b)$. Here we can see that the QCD NLO correction slightly
suppresses the LO differential cross section
$d\sigma_{LO}/dM_{(W^+b)}$.

\par
\begin{figure}
\centering
\includegraphics[scale=0.36]{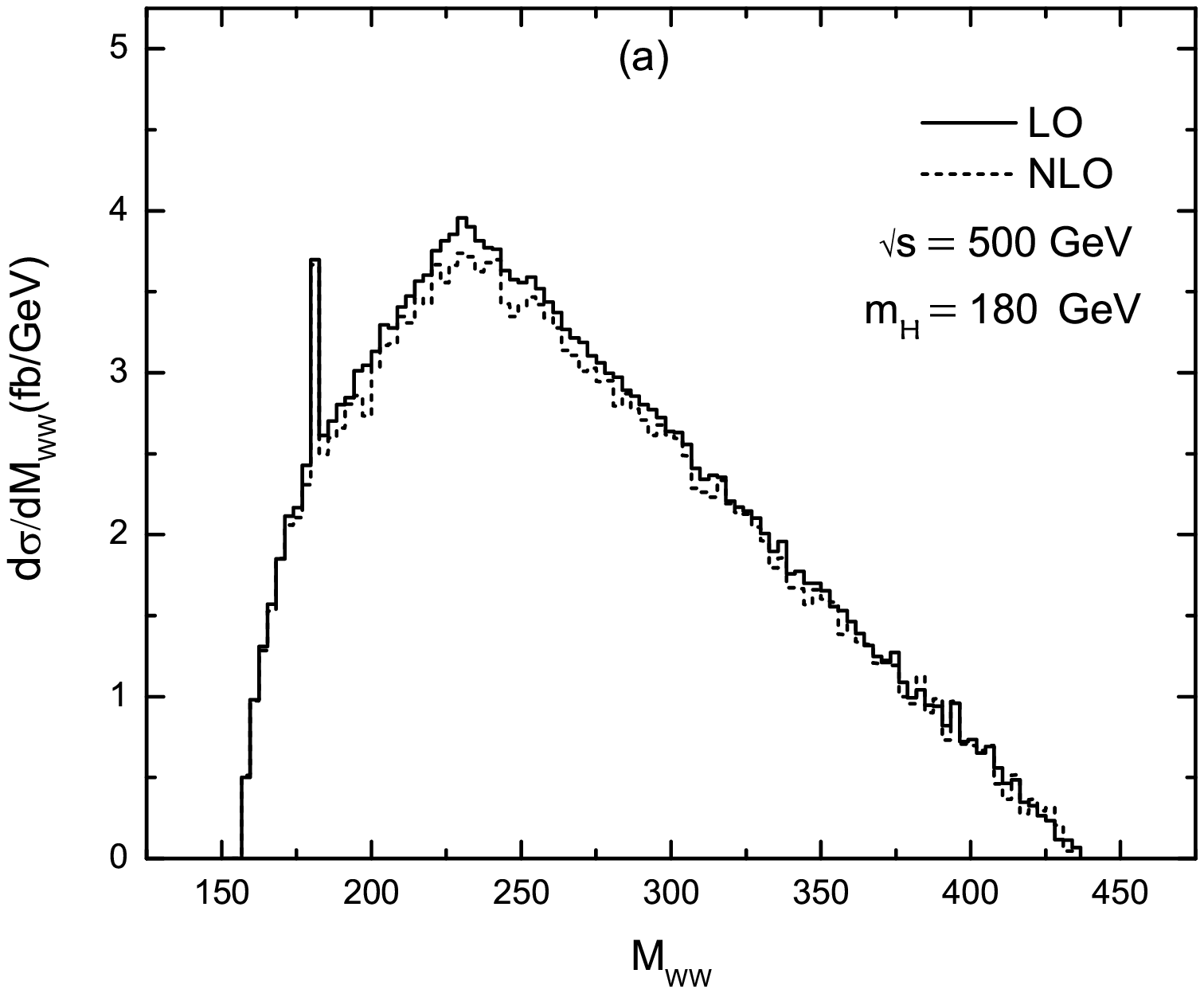}
\includegraphics[scale=0.36]{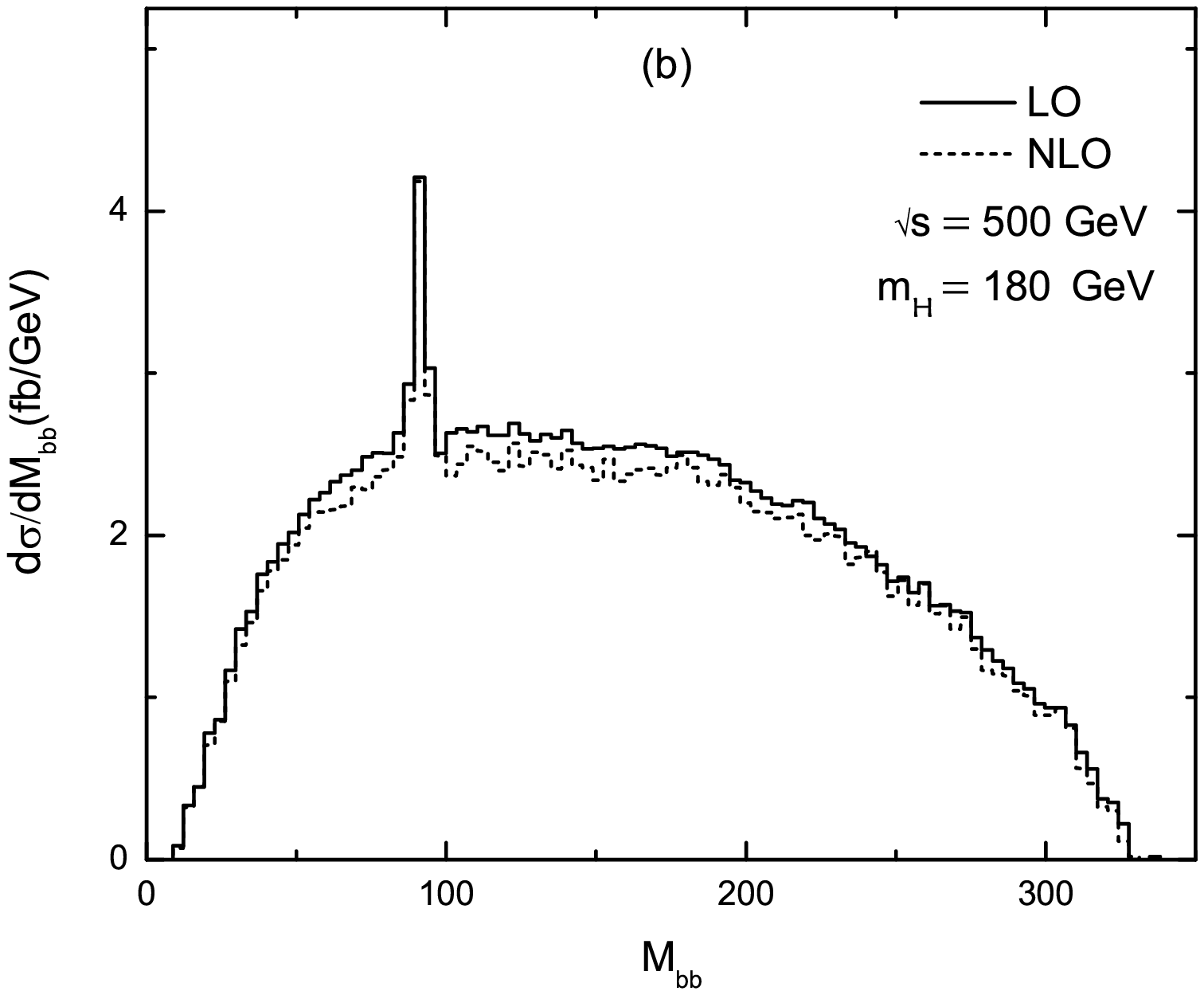}
\caption{\label{fig7} The distributions of the invariant masses of
$(b\bar b)$-pair and $(WW)$-pair with $m_H=180~GeV$ and
$\sqrt{s}=500~GeV$. (a) is for $(WW)$-pair, (b) is for $(b\bar
b)$-pair.}
\end{figure}

As we know, if Higgs boson has a mass larger than $2m_W$, the
$e^+e^- \to H^{0*}Z^{0*} \to W^+W^-b\bar b$ channel will certainly
slightly increase both the LO and QCD NLO corrected cross sections
for $W^+W^-b\bar b$ production due to the Higgs-boson resonant
effect as shown in Fig.\ref{fig4}(a). It will bring a spike on the
distribution of the $(W^+W^-)$-pair invariant mass at the position
of $M_{(WW)}=m_H$. Analogously, the associated real $Z^0$-boson
produced via $e^+e^- \to H^0Z^0$ will induce a spike on the
distribution of the invariant mass $M_{(b\bar b)}$ at the position
of $M_{(b\bar b)}=m_Z$. In Figs.\ref{fig7}(a) and (b) we show the
the distributions of the $WW$- and $b\bar{b}$-pair invariant
masses with $\sqrt{s}=500~GeV$ and $m_H=180~GeV$, respectively. We
can see spikes around the vicinities of $M_{(WW)}\sim m_H\sim
180~GeV$ and $M_{(b\bar b)}\sim m_Z\sim 90~GeV$ in
Fig.\ref{fig7}(a) and Fig.\ref{fig7}(b) respectively, which may be
used to distinguish the $e^+e^- \to H^0Z^0 \to W^+W^-b\bar b$
signature from the corresponding irreducible background \eewwbb.
It shows also the QCD NLO correction obviously modifies the LO
differential cross sections of $d\sigma_{LO}/dM_{(WW)}$ and
$d\sigma_{LO}/dM_{(b\bar b)}$.

\par
Theoretically, the NLO QCD correction to the process $e^+e^-\to
Z^0H^0 \to W^+W^-b\bar b$ with real $Z^0$- and Higgs-boson as
intermediate particles, should be determined only by the NLO QCD
corrections to the $Z^0 \to b\bar b$ decay. As a check to verify
our calculations, we also calculate the correction to the decay
$Z^0\to b\bar b$ with $\sqrt{s}=500~GeV$ and $m_H=180~GeV>2m_W$
and get the K-factors for $e^+e^-\to Z^{0*}H^{0*} \to W^+W^-b\bar
b$ process being 1.0466(1), which is coincident with the result by
calculating the $e^+e^-\to Z^{0*}H^{0*} \to W^+W^-b\bar b$ process
with full QCD NLO diagrams, where the K-factors are 1.046(2).

\vskip 5mm
\section{Summary}
In this paper we calculate the complete one-loop QCD corrections
in the SM to the process \eewwbb at the ILC. We study the
dependence of the LO and QCD NLO corrected cross sections of
process \eewwbb on colliding energy $\sqrt{s}$ and Higgs-boson
mass. We investigate the LO and QCD NLO corrected distributions of
the transverse momenta of final particles and the LO and QCD NLO
corrected differential cross sections of invariant masses of
$Wb$-, $b\bar b$- and $WW$-pair. It shows that NLO QCD correction
obviously modifies the LO cross section of the process \eewwbb,
and when the colliding energy $\sqrt{s}$ goes up from $360~GeV$ to
$1.5~TeV$, the K-factor varies from $1.501$ to $0.847$. The
numerical results show that if $m_H > 2m_W$, the resonant effect
of $H^0$-boson appearing in the $e^+e^- \to H^0Z^0 \to W^+W^-b\bar
b$ channel will induce a little enhancement to the LO and QCD NLO
corrected cross sections for \eewwbb process. We find that it may
be possible to select the $e^+e^- \to H^0Z^0 \to W^+W^-b\bar b$
events from the corresponding irreducible background \eewwbb which
is dominantly produced by the $e^+e^- \to t\bar t \to W^+W^-b\bar
b$ channel by analyzing the invariant masses of final $WW$- and
$b\bar b$-pair.

\vskip 3mm
\par
\noindent{\large\bf Acknowledgments:} This work was supported in
part by the National Natural Science Foundation of China,
Specialized Research Fund for the Doctoral Program of Higher
Education(SRFDP) and a special fund sponsored by Chinese Academy
of Sciences.

\end{document}